\documentclass{jltp}

\usepackage{graphicx} 
\usepackage{amssymb,psfrag,subeqn}

\title{Temperature Dependence of Vortex Nucleation in Gaseous Bose-Einstein Condensates}

\author{Tapio P. Simula, Sami M.~M. Virtanen, and Martti M. Salomaa}

\address{Materials Physics Laboratory, Helsinki University of Technology \\ P.O. Box 2200 (Technical Physics), FIN-02015 HUT, Finland}

\runninghead{T.~P. Simula, S.~M.~M. Virtanen, and M.~M. Salomaa}{Temperature Dependence of Vortex Nucleation in Gaseous BECs}

\newcommand{\beq}{\begin{equation}}
\newcommand{\eeq}{\end{equation}}
\newcommand{\bfr}{\mathbf{r}}
\newcommand{\bfq}{\mathbf{q}}

\newcommand{\Psir}{\hat{\Psi}}
\newcommand{\Psidr}{\hat{\Psi}^\dagger}
\newcommand{\psir}{\hat{\psi}}
\newcommand{\psidr}{\hat{\psi}^\dagger}
\newcommand{\phir}{\phi}
\newcommand{\phicr}{\phi^*}

\newcommand{\ham}{\mathcal{H}}
\newcommand{\Omr}{\mathbf{\Omega}}

\newcommand{\lgl}{\langle}
\newcommand{\rgl}{\rangle}
\newcommand{\rhor}{\tilde{n}}
\newcommand{\delr}{\tilde{m}}
\newcommand{\delcr}{\tilde{m}^*}

\begin{document}

\maketitle

\begin{abstract}
The formation of quantized vortices in trapped, gaseous Bose-Einstein condensates is considered. The thermodynamic stability of vortex states and the essential role of the surface excitations as a route for vortex penetration into the condensate are discussed. Special focus is on finite-temperature \mbox{effects} of the vortex nucleation process. It is concluded that the critical angular frequencies for exciting surface modes with the relevant multipolarities yield, also at finite temperatures, the appropriate thresholds for the nucleation of vortices in dilute Bose-Einstein condensates, in fair agreement with the recent experiments.

\end{abstract}

\section{INTRODUCTION}

The existence of quantized vorticity is an essential part of superfluid phenomena. Quantized flow with $2\pi$ phase winding around a nodal line of a coherent many-body wavefunction is what is usually understood by the concept of a simple vortex line in the framework of quantum mechanics. These topological defects have previously been created and observed, for example, in quantum fluids such as liquid $^4$He and $^3$He, superconductors, and most recently, starting in 1999, in the Bose-Einstein condensates of dilute alkali-atom gases.\cite{Matthews1999b,Madison2000a} 

In type-II superconductors, the order parameter is suppressed along quantized vortex filaments when the external magnetic field exceeds the lower critical field, $H_{c1}$. In higher magnetic fields, the superconducting sample is pierced by an Abrikosov lattice of vortices having normal cores. As the field strength is further increased, the superconducting state vanishes at the upper critical field, $H_{c2}$, as the average distance between the vortices becomes comparable with the size of the vortex core. 

In superfluid helium, vortices are usually generated by mechanically rotating the fluid-containing vessel. Formally, the angular rotation frequency for the helium liquids is the analogue of the magnetic field in superconductors. When superfluid helium is rotated above $\Omega_{c1}$, the fluid is penetrated by quantized vortices. The determination of the exact value of $\Omega_{c1}$ for the helium superfluids is complicated, for example, by effects such as surface roughness of the vessel, remanent vorticity, barrier-tunneling events, and the difficulties in the imaging of the vortex cores. For higher rotation velocities, an increasing number of vortices enter the system forming a regular vortex lattice, but the predicted value for the upper critical frequency $\Omega_{c2}$ is unattainably high and the regime of such extreme rotation velocities is beyond experimental reach.

In dilute Bose-Einstein condensate gases,\cite{Parkins1998a,Dalfovo1999a,Varenna,Courteille2001a,Leggett2001a,Pethick2002a} vortices have been generated by a variety of different methods. The first vortex was created by directly imprinting the characteristic $2\pi$ phase circulation in the two-component condensate.\cite{Matthews1999b} A frequently employed method for creating vortices and vortex lattices---with largest of them containing more than a one hundred vortices---has been the stirring of the condensate by a time-dependent trap anisotropy.\cite{Madison2000a,Madison2000b,Abo2001a,Raman2001a,Haljan2001a,Engels2002a,Hodby2001a} Vortex lattices have also been created through intrinsic rotation, in which case the thermal boson gas is first rotated by a trap anisotropy and the gas is subsequently cooled below the condensation temperature.\cite{Haljan2001a,Engels2002a} In those experiments it is the normal gas surrounding the superfluid that defines the rotating environment for the condensate, instead of the externally rotated trap potential. Since the gaseous condensates are confined by magnetic fields there is no associated surface roughness as in the case of superfluid helium, and thus substantial differences in the nucleation processes are expected to occur between these superfluid systems.

Moreover, vortices have been produced through the snake instability of decaying solitons\cite{Dutton2001a} and in the wake of the turbulent flow generated by a moving object.\cite{Raman2001a,Inouye2001a} Dark solitons have also been seen to decay into vortex rings.\cite{Anderson2001a} Most recently, multiply quantized circulation has been created topologically in dilute Bose-Einstein condensates by adiabatically switching the direction of the magnetic field along the vortex axis.\cite{Leanhardt2002a}

The organization of this paper is as follows: In Sec.~2 we outline the formulation of the mean-field theory relevant for the subsequent discussion. The physical mechanism underlying the vortex-formation process is analyzed both qualitatively and quantitatively at zero and finite temperatures in Sec.~3. Section 4 summarizes our conclusions.

\section{STATIONARY MEAN-FIELD THEORY FORMALISM}
Our starting point is the second-quantized grand-canonical Hamiltonian $\hat{\ham}_\Omega=\hat{\ham}-\Omr\cdot\hat{L}-\mu\hat{N}$ defined in the frame rotating at an angular velocity $\Omr$ by 
\begin{eqnarray}
\hat{\ham}_\Omega &=&\int\Psidr(\bfr) \left[-\frac{\hbar^2}{2m}\nabla^2 + V_{tr}(\bfr)-\mu+i\hbar\Omr\cdot\bfr\times\nabla\right]\Psir(\bfr) \,\mathrm{d}\bfr \nonumber\\&+& \frac{g}{2}\int\Psidr(\bfr)\Psidr(\bfr)\Psir(\bfr)\Psir(\bfr) \,\mathrm{d}\bfr 
\label{Hami}
\end{eqnarray}
where $g$ describes the strength of the two-body interactions between the atoms and $V_{tr}(\bfr)$ denotes the external potential used for spatially confining the atoms.

The bosonic field operator is decomposed in the usual fashion as $\Psir(\bfr)=\phir(\bfr)+\psir(\bfr)$, where $\phir(\bfr)$ stands for the macroscopic wavefunction of the condensate. The mean-field scheme corresponds to approximating the cubic and quartic operator products according to\cite{Griffin1996a}
\begin{eqnarray}
\psidr(\bfr)\psidr(\bfr)\psir(\bfr)\psir(\bfr) &\approx& 4\rhor(\bfr)\psidr(\bfr)\psir(\bfr) +\delr(\bfr)\psidr(\bfr)\psidr(\bfr)+\delcr(\bfr)\psir(\bfr)\psir(\bfr) \nonumber \\
\psidr(\bfr)\psir(\bfr)\psir(\bfr) &\approx& \delr(\bfr)\psidr(\bfr)+2\rhor(\bfr)\psir(\bfr),
\label{MFa}
\end{eqnarray}
where we have introduced the noncondensate density $\rhor(\bfr) \equiv \lgl\psidr(\bfr)\psir(\bfr)\rgl$ and the anomalous correlation function $\delr(\bfr) \equiv \lgl\psir(\bfr)\psir(\bfr)\rgl$ describing the condensate-induced correlations between the noncondensate particles.

Inserting the Bogoliubov decomposition for the field operator and the mean-field approximations in Eq.~(\ref{Hami}), we obtain an effective quadratic Hamiltonian, $\hat{\ham}_{eff}$. The linear terms in $\hat{\ham}_{eff}$ are eliminated by the stationary generalized Gross-Pitaevskii equation, 
\beq
\left[\mathcal{L}-g|\phir(\bfr)|^2\right]\phir(\bfr) + g\delr(\bfr)\phicr (\bfr)= 0,
\label{GP}
\eeq
for the macroscopic wavefunction $\phi(\bfr)$ and the chemical potential $\mu$. Above,
\beq
\mathcal{L}\equiv-\frac{\hbar^2}{2m}\nabla^2 + V_{tr}(\bfr)-\mu+2g\left[|\phir(\bfr)|^2 + \rhor(\bfr)\right] +i\hbar\Omr\cdot\bfr\times\nabla,
\eeq
and the condensate wavefunction is normalized to the fixed number of particles, $N$, according to $\int\left[|\phi(\bfr)|^2+\tilde{n}(\bfr)\right]\,\mathrm{d}\bfr=N$.

The remaining terms in the effective Hamiltonian may be diagonalized by using the canonical Bogoliubov transformation 
\beq
\psir(\bfr)=\sum_\bfq \left[ u_\bfq(\bfr)\eta_\bfq-v^*_\bfq(\bfr)\eta^\dagger_\bfq\right],
\label{Transform}
\eeq
and demanding the quasiparticle wavefunction amplitudes $u_\bfq(\bfr)$ and $v_\bfq(\bfr)$ to obey the coupled Hartree-Fock-Bogoliubov eigenvalue equations 
\begin{subequations}
\begin{eqnarray}
\mathcal{L} u_\mathbf{q}(\bfr)- \mathcal{M}v_\mathbf{q}(\bfr)&=&E_\mathbf{q}u_\mathbf{q}(\bfr) \\
\mathcal{L}^*v_\mathbf{q}(\bfr)- \mathcal{M}^* u_\mathbf{q}(\bfr)&=&-E_\mathbf{q}v_\mathbf{q}(\bfr)  
\end{eqnarray}
\label{Bogo}
\end{subequations}
where $\mathcal{M}\equiv g[\delr(\bfr)+\phi^2(\bfr)]$ and $E_q$ label the quasiparticle eigenenergies.
Moreover, conservation of the bosonic commutation relations implies the normalization condition
\beq
\int[u^*_i(\bfr)u_j(\bfr)-v_i^*(\bfr)v_j(\bfr)]\,\mathrm{d}\bfr =\delta_{ij}
\label{norm}
\eeq
for the quasiparticle amplitudes. The positive-norm quasiparticle states with negative eigenenergies $E_q$ are referred to as the `anomalous modes'. 

Within the mean-field approximation, the quasiparticles are noninteracting, the corresponding creation and annihilation operators obeying the following equilibrium relations;
\begin{subequations}
\begin{eqnarray}
&&[\eta_i,\eta^\dagger_j]=\delta_{ij} \\
&&\lgl\eta_i\eta_j\rgl=\lgl\eta^\dagger_i\eta^\dagger_j\rgl=0 \\
&&\lgl\eta^\dagger_i\eta_j\rgl=\delta_{ij}n_i\equiv\delta_{ij}(e^{E_i/k_{B}T}-1)^{-1},
\end{eqnarray}
\label{quasicles}
\end{subequations}
where $n_i$ is the Bose-Einstein distribution function. Using these relationships, the self-consistent mean fields may be expressed in the form;
\begin{subequations}
\begin{eqnarray}
\rhor(\bfr) &=& \sum_{i} \left[n_i\left(|u_i(\bfr)|^2+|v_i(\bfr)|^2\right)+|v_i(\bfr)|^2\right], \\
\delr(\bfr) &=& -\sum_{i}\left[2n_iu_i(\bfr)v_i^*(\bfr)+u_i(\bfr)v_i^*(\bfr)\right].
\end{eqnarray}
\label{MFb}
\end{subequations}
Equations (\ref{GP}), (\ref{Bogo}) and (\ref{MFb}) constitute a self-consistent set of equations whose solution for a partially Bose-Einstein condensed system of atoms require iterative methods.

The expectation value of the effective Hamiltonian $\hat{\ham}_{eff}$ is found to be
\begin{eqnarray}
\lgl\hat{\ham}_{eff}\rgl&=&\int\phicr(\bfr)\left[\mathcal{L}-g\frac{3}{2}|\phir(\bfr)|^2-2g\rhor(\bfr)\right]\phir(\bfr)  \,\mathrm{d}\bfr  \nonumber \\
&+& \frac{g}{2}\int \left[2|\delr(\bfr)|^2+\phi^2(\bfr)\delcr(\bfr) +\phi^{*^2}(\bfr)\delr(\bfr)\right] \,\mathrm{d}\bfr \nonumber \\
&+& \int\lgl\psidr(\bfr)\mathcal{L}\psir(\bfr)\rgl \,\mathrm{d}\bfr. 
\label{Effham}
\end{eqnarray}
Taking Eqs.~(\ref{Transform}) and (\ref{quasicles}) into account, the last line in Eq.(\ref{Effham}) is expanded in the quasiparticle basis as
\beq
\lgl\psidr(\bfr) \mathcal{L} \psir(\bfr)\rgl= \sum_{i} \left[n_i u^*_i(\bfr) \mathcal{L} u_i(\bfr)+(n_i+1) v_i(\bfr) \mathcal{L} v^*_i(\bfr)\right]. 
\label{eq1}
\eeq
Multiplying Eq.~(\ref{Bogo}a) and the complex conjugate of Eq.~(\ref{Bogo}b) by $u_i(\bfr)$ and $v_i(\bfr)$, respectively, we find   
\begin{eqnarray}
\lgl\psidr(\bfr) \mathcal{L} \psir(\bfr)\rgl&=& \sum_{i}\left[ n_i(\mathcal{M}u_i^*(\bfr)v_i(\bfr) + |u_i(\bfr)|^2E_i)\right. \nonumber\\ 
&+& \left.(n_i+1)( \mathcal{M}v_i(\bfr)u_i^*(\bfr) - |v_i(\bfr)|^2E_i) \right].\nonumber \\
\end{eqnarray}
Within the framework of the mean-field approximation, Eqs.~(\ref{MFb}), the above expression reduces into
\beq
\lgl\psidr(\bfr) \mathcal{L}\psir(\bfr)\rgl=\sum_{i}\left[n_i(|u_i(\bfr)|^2-|v_i(\bfr)|^2)-|v_i(\bfr)|^2\right]E_i -\mathcal{M}\tilde{m}^*(\bfr).
\eeq 
The quasiparticle normalization condition, Eq.~(\ref{norm}), thus results after rearrangement of the terms into
\begin{eqnarray}
\lgl\hat{\ham}_{eff}\rgl&=&\int\phicr(\bfr)\left[\mathcal{L}-g\frac{3}{2}|\phir|^2-2g\rhor(\bfr)\right]\phir(\bfr)  \,\mathrm{d}\bfr  \nonumber \\
&+& \frac{g}{2}\int \left[
 \phi^{*^2}(\bfr)\delr(\bfr) - \phi^2(\bfr)\delcr(\bfr)
\right] \,\mathrm{d}\bfr \nonumber \\
&+&\sum_{i}n_iE_i-\sum_{i}E_i\int|v_i(\bfr)|^2 \,\mathrm{d}\bfr. 
\label{Expham}
\end{eqnarray}
The second line of Eq.~(\ref{Expham}) vanishes since, by definition, $\lgl\hat{\ham}_{eff}\rgl$ is a real function, thereby Im$\int\tilde{m}^*(\bfr)\phi^2(\bfr)\,\mathrm{d}\bfr =0$, and we find
\begin{eqnarray}
\lgl\hat{\ham}_{eff}\rgl&=&-\frac{g}{2}\int|\phir(\bfr)|^4 \,\mathrm{d}\bfr  -2 g\int|\phir(\bfr)|^2\rhor(\bfr) \,\mathrm{d}\bfr   - g\int\tilde{m}(\bfr)\phi^{*2}(\bfr)\,\mathrm{d}\bfr \nonumber\\
&+& \sum_{i}n_iE_i-\sum_{i}E_i\int|v_i(\bfr)|^2 \,\mathrm{d}\bfr 
\end{eqnarray}
where the first line is expressed using the Gross-Pitaevskii equation.
The entropy of the noncondensate atoms equals 
\beq
S=-k_B\sum_{i}[n_i\ln n_i -(1+n_i)\ln(1+n_i)]
\label{entropy}
\eeq
while the free energy of the gas is finally obtained from 
\beq
F=\lgl\hat{\ham}_{eff}\rgl+\mu N -TS.
\label{free}
\eeq

\section{MECHANISM FOR VORTEX FORMATION}

In a typical vortex-creation experiment, the Bose-Einstein condensate is rotated at a certain angular velocity by the time-dependent trap anisotropy. After the stirring, the condensate is allowed to equilibrate. Subsequently, the trap is turned off and the cloud is allowed to expand such that the possible presence of the vortex cores can be imaged. Computationally, vortices have been verified to enter the condensate from its boundary region by a number of simulations.\cite{Butts1999a,Caradoc-Davies1999a,Tsubota2002a,Penckwitt2002a}

Characteristic for these rotating-bucket experiments has been the observation that the vortices are only seen at substantially higher angular frequencies than what is expected from the basis of purely thermodynamic equilibrium arguments. The reported angular rotation frequency thresholds for vortex nucleation approximately include 0.1,\cite{Raman2001a} 0.3,\cite{Abo2001a} 0.4,\cite{Haljan2001a} and 0.7\cite{Hodby2001a,Chevy2000a,Madison2001a} in the units of the transverse trap frequency. A number of theoretical investigations have been performed to explain these observations.\cite{Isoshima1999a,Feder1999a,Dalfovo2000a,Feder2001a,Garcia-Ripoll2001a,Mitsubishi,Garcia-Ripoll2001b,Sinha2001a,Anglin2001a,Williams2001a,Anglin2002a,Simula2002b,Garcia-Ripollpreprint,Muryshevpreprint,Kramerpreprint}

The relatively high nucleation thresholds may be qualitatively understood by considering the energy difference between the vortex and nonvortex states as a function of the position of the vortex in the condensate;\cite{Fetterreview} in a trap rotating at the thermodynamic equilibrium frequency to support an axisymmetric vortex line, this self-energy of the vortex features a maximum on the condensate surface which prevents the vortex from penetrating the system. Thus much higher stirring frequencies are needed in order to remove the energy barrier at the condensate surface and hysteretic behavior results.

In the following, we consider singly quantized, axisymmetric vortex \mbox{states} radially confined by a harmonic trapping potential with the frequency $\omega_\perp$ perpendicular to the rotation axis. Although the periodic boundary conditions applied in the axial direction makes a direct comparison with a specific experimental setup difficult, the relevant physics described by this system closest resembles that of oblate/spherical condensate geometries in which, for instance, vortex bending is negligible. The axisymmetric condensate states are obtained from the Gross-Pitaevskii equation and the Bogoliubov Eqs.~(\ref{Bogo}) are solved for the quasiparticle excitations which serve to provide information on the local energetic stabilities of the states. The thermodynamic critical angular frequency for vortex stability is determined by the condition that the free energies, see Eq.~(\ref{free}), of the vortex and the nonvortex states become equal in the rotating frame of reference in which the Hamiltonian is time-independent. For technical details of the numerical computations performed, see Ref.~\onlinecite{Simula2002b}.

\subsection{Local Energetic Stability}

\begin{figure}[!b]
\psfrag{x}[c][]{\small \raisebox{-3ex}{$l$}}
\psfrag{y}[c][]{\small \raisebox{2ex}{$E_q [\hbar\omega_\perp]$}}
\centering
\begin{minipage}{.48\linewidth}
\includegraphics[width=\linewidth]{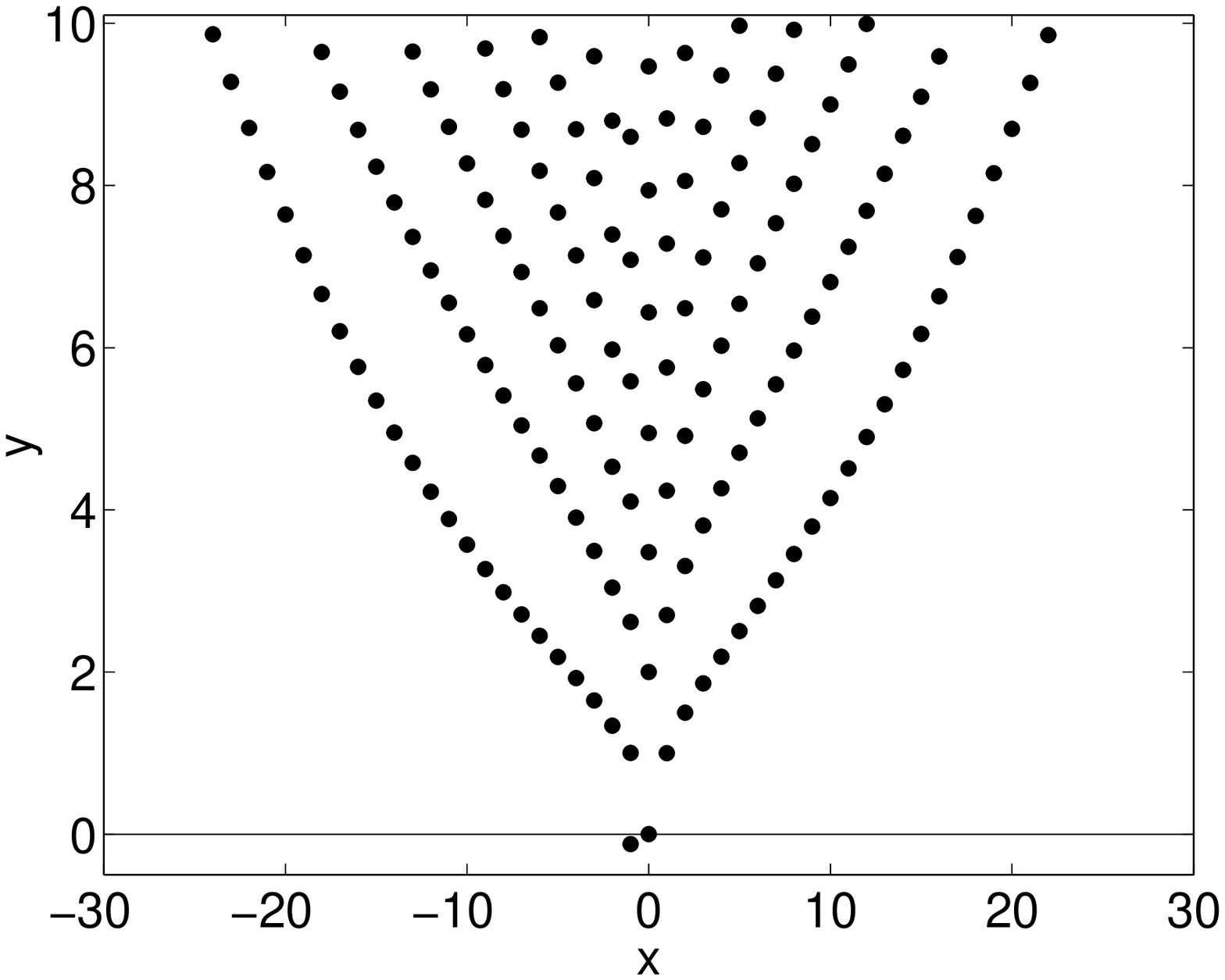}
\end{minipage}
\begin{minipage}{.48\linewidth}
\includegraphics[width=\linewidth]{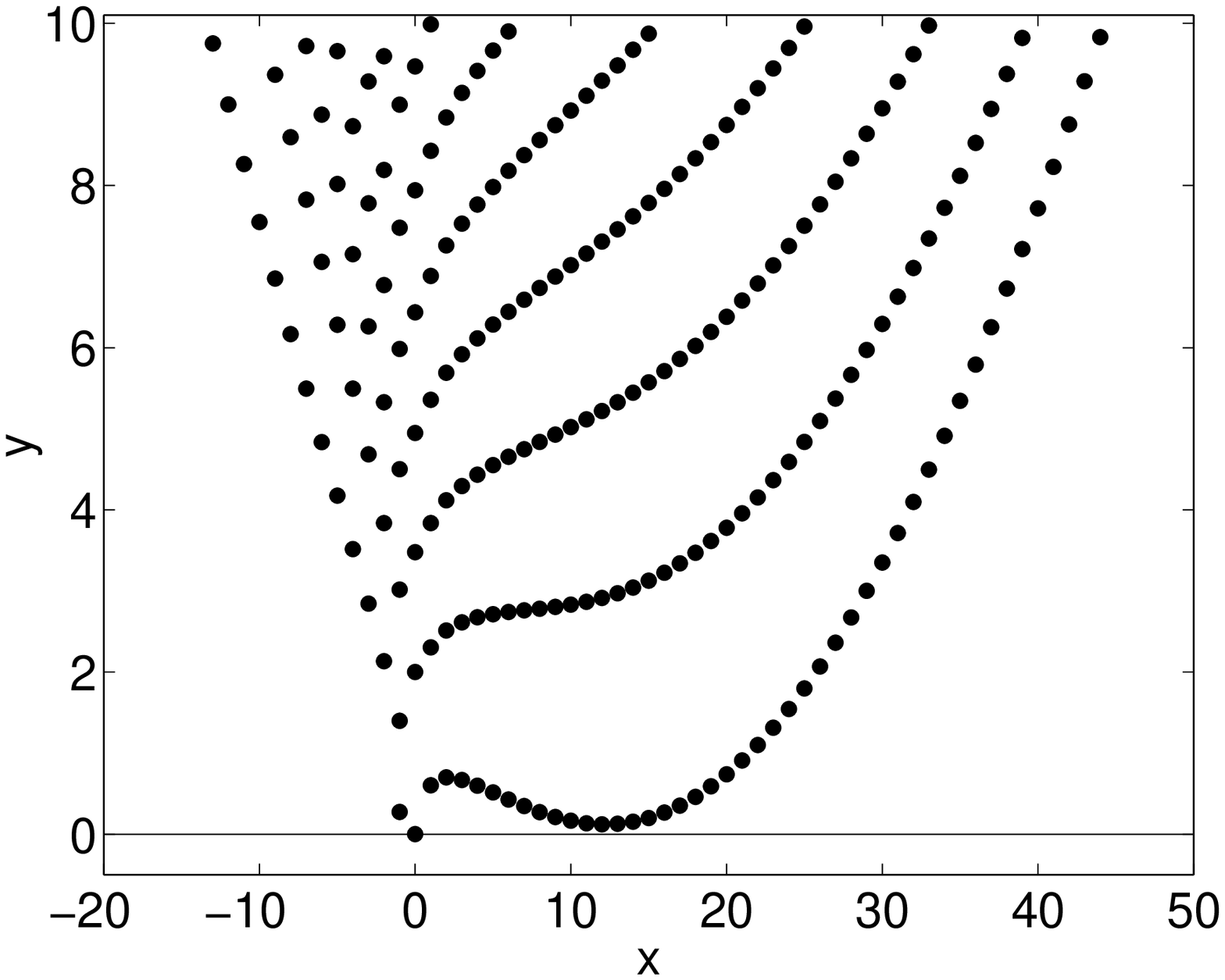}
\end{minipage}
\begin{picture}(0,0)
\put(-336,55){\scriptsize (a)}
\put(-159,55){\scriptsize (b)}
\end{picture}
\caption{Computed windows of the Bogoliubov quasiparticle excitation spectra for a singly quantized vortex state as functions of the angular momentum quantum number, $l$. (a) The spectrum of the nonrotating system exhibits one anomalous vortex core localized mode for $l=-1$ with a negative energy with respect to the condensate state. Such an anomalous mode also corresponds to the precessing motion of an off-centered vortex line and implies local energetic instability of the state.\cite{Feder2001a,Svidzinsky2000a} The suggested possible decay channel of such a state in the presence of dissipation is the outward spiraling motion of the vortex leading to vortex annihilation on the condensate surface.\cite{Rokhsar1997a} (b) The corresponding spectrum computed in the frame rotating at an angular frequency close to $\Omega_\ell$, see Eq.~(\ref{Landau}). The anomalous vortex core mode ($l=-1$) is Doppler-shifted to positive energy but the lowest surface modes now osculate the condensate energy level. Further increase of the rotation would allow for the generation of anomalous surface modes, implying an instability of the singly quantized vortex state to a state of an increasing number of vortices.}\label{Fig3}
\end{figure}
In nonrotating traps, according to zero-temperature Bogoliubov theory, an axisymmetric vortex state is locally unstable in the sense that the vortex would spiral out of the condensate if the dissipation in the system is significant.\cite{Rokhsar1997a} In other words, the self-energy of the vortex decreases monotonically as the radial distance of the vortex from the trap center increases.\cite{Fetterreview} Such local energetic instability is also manifest in the spectrum of the vortex state which contains anomalous mode(s) localized in the vortex core, see Fig.~\ref{Fig3}a. These elementary excitations possess negative energies with \mbox{respect} to the condensate ground state and thus the condensate could lower its energy by transferring particles into the anomalous mode.

However, when the trap rotation frequency is increased, a plateau in the self-energy is formed corresponding to the onset of metastability of the axisymmetric vortex state.\cite{Fetterreview} In general, the local energetic stability of the condensate state may be identified with the positivity of the quasiparticle excitation spectrum of the system. In terms of the Bogoliubov excitation spectrum, the local (meta)stability frequency $\Omega_{m}$ corresponds to the frequency $\omega_{a}$ of the lowest anomalous mode which acquires an energy $\hbar\Omega$ in the frame rotating at the angular frequency $\Omega$. 

\subsection{Thermodynamic Equilibrium}

As the rotation frequency of the trap exceeds the  metastability frequency $\Omega_{m}$, a minimum develops in the self-energy function at the trap center. At a sufficiently high frequency, $\Omega_c$, the free energies of the vortex and the nonvortex states evaluated in the rotating frame of reference coincide on the trap axis, indicating the onset of thermodynamic stability of an axisymmetric vortex. At this stage, on the condensate boundary there still remains the energy barrier preventing the vortex entry which has to be overcome. This implies hysteretic behavior when the rotation frequency is ramped above the nucleation threshold and back again. 

The (global) thermodynamic stability implies local energetic stability of the system. However, in elongated trap geometries the critical frequency for the local stability of an axisymmetric vortex line may even exceed the corresponding thermodynamic value $\Omega_c$.\cite{Feder2001a} This is because the true equilibrium configuration in prolate trap geometries corresponds to a bent vortex line.\cite{Garcia-Ripoll2001b,Rosenbusch2002b} The additional rotation required for the local stabilization of an axisymmetric vortex line is consumed in the straightening of the otherwise bent vortex.  

\subsection{Landau Criterion}

In order to explain the high values of the observed vortex nucleation thresholds, it has been argued that the correct condition for the removal of the energetic barrier on the condensate surface leading to the vortex formation through the surface instability is given by the Landau criterion for the generation of quasiparticle excitations in the system.\cite{Anglin2002a} Specializing to a rotationally invariant systems, the Landau critical angular frequency for the creation of the surface modes in the condensate is given by
\beq
\Omega_\ell=\min_l\left\{ \frac{\omega_l}{\l}\right\},
\label{Landau}
\eeq
where $\hbar\omega_l$ is the energy of a quasiparticle carrying the angular momentum $\hbar l$. These surface modes are quasiparticle excitations corresponding to revolving density perturbations localized around the condensate surface. Again, in terms of the excitation spectrum, the Landau critical frequency corresponds to the rotation frequency at which there emerges anomalous negative-energy modes localized on the condensate surface in the elementary excitation spectrum of the system, see Fig.~\ref{Fig3}b. 

\begin{figure}[!t]
\psfrag{x}[c][]{\small \raisebox{0ex}{$l$}}
\psfrag{y}[c][]{\small \raisebox{3ex}{$\omega_l/l\;[\omega_\perp]$}}
\centerline{\includegraphics[height=2.5in]{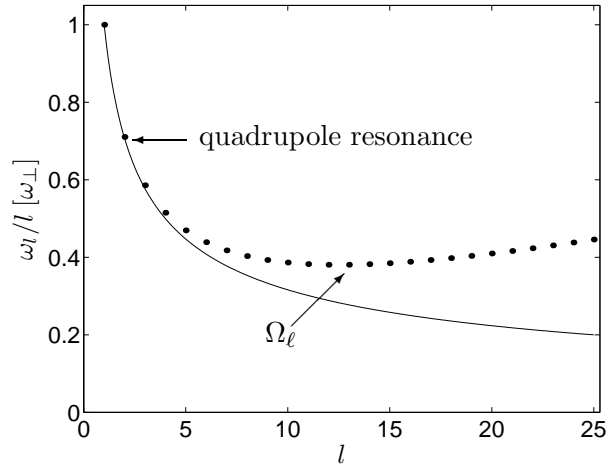}}
\begin{picture}(0,0)
\put(130,136.5){\vector(-1,0){20}} 
\put(135,134.5){quadrupole resonance}
\put(170,66){\vector(1,1){20}}
\put(160,60){$\Omega_\ell$}
\end{picture}
\caption{Energies per angular momentum for surface modes as functions of the angular-momentum quantum number $l$ calculated within the zero-temperature Bogoliubov approximation. The solid line is the hydrodynamic prediction $1/\sqrt{l}$ yielding a fair approximation for the lowest multipole modes.\cite{Dalfovo2000a} For higher values of angular momenta, the deviation becomes significant as the hydrodynamic value for $\omega_l/l$ diverges whereas the Bogoliubov spectrum exhibits a minimum corresponding to the Landau critical angular frequency, $\Omega_\ell$. The quadrupole resonance is expected to occur at $\Omega\approx 0.7 \omega_\perp$ in agreement with the experimental observations in which the condensate has been stirred by a quadrupole drive.}  
\label{Fig1}
\end{figure}

However, in a number of recent experiments, the emergence of the vortices has been seen only for yet higher rotation velocities of the trap. According to the current understanding, those values are best explained in terms of the excitation of the quadrupolar surface modes. In Ref.~\onlinecite{Raman2001a}, patterns of laser beams with different multipolarities were used and the vortex nucleation resonances were observed for corresponding frequencies for the excitation of multipolar surface modes. Thus it is understood that if the rotating anisotropy resonantly excites certain surface modes, the vortex nucleation threshold is determined by the frequency of those excitations rather than by the minimum of $\omega_l/l$.

In the hydrodynamic limit, the dispersion relation for the collective quasiparticle excitations of a trapped Bose-Einstein condensate was obtained in Ref.~\onlinecite{Stringari1996a}. For surface modes it reduces to $\omega_l=\sqrt{l}\omega_\perp$ which is expected to be accurate for the lowest angular-momentum states, see Fig.~\ref{Fig1}. Especially, for the quadrupole mode it yields $\omega_l/l\approx 0.7\,\omega_\perp$ in reasonable agreement with the observations reported in Refs.~\onlinecite{Hodby2001a,Chevy2000a,Madison2001a}. Although the spectrum, and especially $\Omega_\ell$, depends on the shape of the trapping potential, the value for the quadrupole resonance $l=2$ is only slightly altered under a change of the trap geometry.\cite{Dalfovo2000a}

\begin{figure}[!t]
\psfrag{x}[c][]{\small \raisebox{0ex}{$r [\mu{\rm m}]$}}
\psfrag{y}[c][]{\small \raisebox{3ex}{$\rho [\mu{\rm m}^{-3}]$}}
\centerline{\includegraphics[height=2.5in]{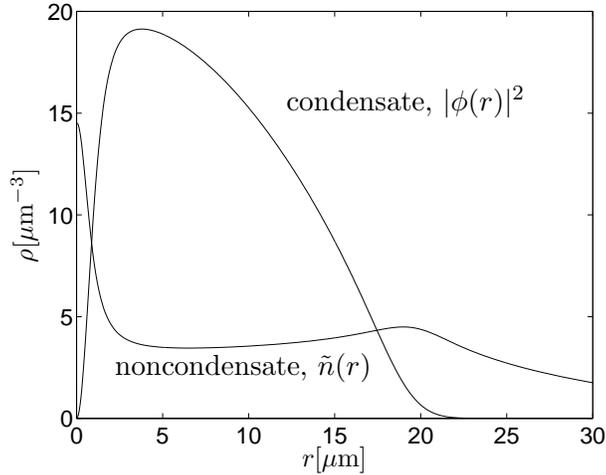}}
\begin{picture}(0,0)
\put(105,50){noncondensate, $\tilde{n}(r)$}
\put(170,150){condensate, $|\phi(r)|^2$}
\end{picture}
\caption{Radial density distributions of the condensate and the noncondensate components for $T\approx T_c/2$. For the sake of clarity, the noncondensate density is scaled up by a factor of 5. The condensate density vanishes on the vortex axis where the noncondensate accumulates forming an additional self-stabilizing potential for the vortex state. On the boundary of the condensate, the noncondensate bulges due to the mutual repulsion between the two components.}
\label{Fig0}
\end{figure}

Notice also that the thermodynamic critical frequency $\Omega_c$ may be interpreted as the Landau criterion $\Omega_c=\Delta F/L$ for the generation of a vortex excitation. Here $\Delta F=F(\mathrm{vortex})-F(\mathrm{nonvortex})$ is the `activation energy' needed for the creation of the vortex and $L=N\hbar$ is its angular momentum. However, the difference between $\Omega_c$ and $\Omega_\ell$ is that the latter is the frequency needed for generating a single excitation whereas the former corresponds to exciting all the particles into the unit angular-momentum state simultaneously.

\subsection{Critical Frequencies at Finite Temperatures}
A characteristic feature for all the self-consistent, first-order finite-tem\-perature theories considered for vortex states in Refs.~\onlinecite{Isoshima1999b} and \onlinecite{Virtanen2001a} is that none of them contain anomalous vortex core modes in distinction to the prediction of the zero-temperature Bogoliubov approximation. This is because of the core-filling effect of the noncondensate component lifting the anomalous modes to positive energies, see Fig.~\ref{Fig0}. Moreover, it suggests that the axisymmetric vortex state could be metastable even in the nonrotating trap. However, these approximations neglect the noncondensate dynamics causing shifts in the energies of the quasiparticle modes. For a further discussion on thermal effects on the local stability of vortex states see, for example, Ref.~\onlinecite{Virtanenpreprint} and the references therein.

\begin{figure}[!ht]
\psfrag{x}[c][]{\small \raisebox{-3ex}{$T$ [n$K$]}}
\psfrag{y}[c][]{\small \raisebox{3ex}{$\Omega\;[\omega_\perp]$}}
\centerline{\includegraphics[height=2.5in]{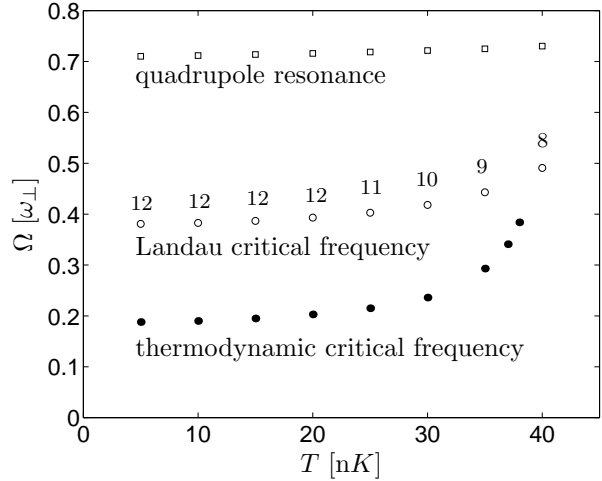}}
\begin{picture}(0,0)
\put(114,160){\small quadrupole resonance}
\put(266,136){{\scriptsize 8  }}
\put(243,126){{\scriptsize 9  }}
\put(220,121){{\scriptsize 10 }}
\put(199,118){{\scriptsize 11 }}
\put(178,115){{\scriptsize 12 }}
\put(156,114){{\scriptsize 12 }}
\put(134,113){{\scriptsize 12 }}
\put(112,112){{\scriptsize 12 }}
\put(114,95){\small Landau critical frequency}
\put(114,57){\small thermodynamic critical frequency}
\end{picture}
\caption{Thermodynamic ($\bullet$) and Landau ($\circ$) critical angular frequencies, and the quadrupole resonance ({\tiny$\blacksquare$}) for vortex formation as functions of temperature. All of these frequencies increase monotonically for increasing temperature but the temperature dependence is  significant only at higher temperatures for which the validity of the Popov approximation not treating the dynamics of the noncondensate is questionable. The numbers above the circles denote the corresponding multipolarity $l$ of the destabilizing surface mode. The critical temperature for the system is given by\cite{Stringari1999a} $T_c(\Omega)=T_c^0(1-\Omega^2/\omega_\perp^2)^{1/3}$ where $T_c^0\approx 45$ nK for the parameter values used.}  
\label{Fig2}
\end{figure}

In Fig.~\ref{Fig2} we have plotted the thermodynamic critical frequency ($\bullet$), the critical frequency given by the Landau criterion ($\circ$) for generation of surface excitations, and the quadrupole resonance ({\tiny$\blacksquare$}) as functions of temperature. The numbers above the squares denote the multipolarity $l$ of the destabilizing surface mode. The decrease in the multipolarity as temperature is increased is caused by the shrinking of the condensate with increasing temperature. All of the critical frequencies remain fairly constant for temperatures below 0.6 $T_c$ where the Popov approximation neglecting the off-diagonal mean-fields $\tilde{m}(\bfr)$ is expected to be in reasonable agreement with the experiments. However, even for higher temperatures the variation of these frequencies is such that the critical frequency thresholds for the formation of vortices is rather determined by external parameters, such as the geometry of the confining trap and the shape of the rotating perturbation driving the condensate towards the vortex state, than the temperature of the system.\cite{Simula2002b} The results for the nucleation thresholds at finite temperatures, see Fig.~\ref{Fig2}, are in fair accordance with the experimentally observed values, see Sec.~3, and with those predicted by the zero-temperature field theories.\cite{Dalfovo2000a} The quadrupole resonance occurs at $\approx 0.7\,\omega_\perp$, and for the temperatures accessible in the experiments with oblate/spherical condensate shapes for which our computational results should closest be applicable, the Landau critical frequency yields values close to $0.3\,\omega_\perp - 0.4\,\omega_\perp$. 

\section{DISCUSSION}

The formation process of vortex lattices in a stirred Bose-Einstein condensate studied at MIT exhibited only a weak dependence on temperature.\cite{Abo-Shaeer2002a} Contrary to this, the experiment performed at ENS reported a qualitative temperature dependence in the lattice ordering.\cite{Madison2000a} In the light of these experiments and the recent pronounced interest in the vortex formation phenomena in superfluids, it is instructive to investigate theoretically the finite-temperature effects on the vortex nucleation in weakly interacting, gaseous Bose-Einstein condensates. 

We have computed the thermodynamic critical frequency $\Omega_c$ for the stability of an axisymmetric vortex line as well as the resonance frequencies of the surface modes for a range of temperatures providing, for instance, the Landau critical frequency $\Omega_\ell$ for vortex nucleation. Only a weak temperature dependence in the critical rotation frequencies were found in the temperature range within which the Popov approximation applied is expected to yield reliable predictions. Thus the conclusion is that the high observed values for the vortex nucleation frequency are not solely related to the noncondensate gas.     

It is understood that the lower critical trap rotation frequency $\Omega_{c1}$ for the formation of vortices in gaseous Bose-Einstein condensates is rather given by the Landau criterion $\Omega_\ell$ for the generation of surface excitations than by the thermodynamic value $\Omega_c$ for the stability of an axisymmetric vortex. If the rotating drive of the condensate possesses a well-defined multipolarity $l$, it is the corresponding surface mode which determines the threshold nucleation frequency, instead of $\Omega_\ell$. Moreover, the low stirring frequency reported in Ref.~\onlinecite{Raman2001a} is probably explained by the fact that the actual condensate flow velocity around the small stirrer beam employed is likely to exceed the rotational velocity of the beam, hence yielding a much higher effective rotation frequency.\cite{Anglin2001a}

In contrast to the helium superfluids, the limit of an upper critical rotation, $\Omega_{c2}$, should be within experimental reach in the gaseous Bose-Einstein condensates. Recently, such condensates in rapidly rotating traps have attracted much interest.\cite{Ho2001a,Fetter2001a,Recati2001a,Fischer2002a} When the rotation frequency $\Omega$ of the trap reaches the harmonic trapping frequency $\omega_\perp$, the confinement of the atoms is lost and they leave the trap.\cite{Rosenbusch2002a} Close to this limit, the system may enter the quantum-Hall-like regime\cite{Wilkin2000a,Viefers2000a,Paredes2001a} and the vortex cores are predicted to melt into a single giant multiply quantized vortex.\cite{Fischer2002a,Kasamatsu2002a,Lundh2002a} If the atoms could be held together above $\omega_\perp$, for instance, by additional stronger than harmonic potentials, the physics in the vicinity of the upper critical frequency could be experimentally explored. Also the possibility of a Kosterlitz-Thouless transition at such extreme rotational conditions has been suggested.\cite{Fischer2002a} 

In addition to the giant vortices in the rapidly rotating traps, a generation of multiply quantized vorticity in dilute Bose-Einstein condensates have been suggested by applying additional pinning potentials for stabilizing the multiquantum vortex states.\cite{Simula2002a} Moreover, vortices with circulation in multiplets of two have recently been created\cite{Leanhardt2002a} by continuously changing the direction of the axial magnetic field, as first suggested theoretically.\cite{Isoshima2000a,Ogawa2002a}

\section*{ACKNOWLEDGMENTS}
We thank the Center for Scientific Computing for computer resources,
and the Academy of Finland and the Graduate School in Technical Physics
for support. One of us (TPS) is grateful to the Research Council of Helsinki University of Technology for a postgraduate scholarship. Discussions with M. Nakahara, T. Isoshima, and M. M\"ott\"onen are appreciated. The organizers and participants of the Kevo Winter School are thanked for the pleasant atmosphere during the week at the Kevo Biological Station.

\end{document}